\documentstyle[prl,aps,preprint]{revtex}
\begin{document}
\begin{center}
   {\Large \bf The hole-hole superconducting pairing in $t$-$J$ model
induced by the long-range spin-wave exchange}
\end{center}
V.V.Flambaum$^1$, M.Yu.Kuchiev$^{1,2}$, and O.P.Sushkov$^{1,3}$\\
{\em $^1$ School of Physics, The University of New South Wales\\
   P.O.Box 1, Kensington, NSW 2033, Australia\\
$^2$ A.F.Ioffe Physical-Technical Institute, 194021 St. Petersburg, Russia\\
$^3$Budker Institute of Nuclear Physics, 630090 Novosibirsk, Russia}

\begin{abstract}
Long-range $(r \sim p_F^{-1} \gg lattice \ spacing)$ spin-wave exchange
produces a very strong pairing of the
holes. The different symmetry solutions of BCS-type equation for the
superconducting gap $\Delta$ are found. The most strong pairing
corresponds to $d$-wave symmetry. The physical reasons for the
strong pairing are: 1)large velocity of the spin-wave,
2) a ``collapse'' effect in the attractive potential ($-1/r^2$)
describing the interaction between two holes induced by spin-wave
exchange.
3)strongly asymmetric hole dispersion.
At concentration of the holes $\delta \sim 0.01-0.3$ we get
$\Delta \sim T_c \sim \epsilon_F$. At very low concentration of
the holes the many-body wave function of the ground state
decays into the product of two-hole bound states.
\end{abstract}

\newpage
%****************************************************************************
\section{Introduction}
%****************************************************************************
Many authors believe that antiferromagnetic spin fluctuations play an
important role in the formation of superconducting state in high-$T_c$
superconductors Refs.\cite{Sch9,Kam1,Kam2,Bul1,Bul2,Mon1,Mon2,Dah3}.
Unfortunately these approaches are based on rather arbitrary
suppositions about the interaction and in our opinion do not
take into account the main features of spin-wave dynamics in the
system. We believe that now the level of understanding of $t-J$
model allows one to take into account these features.

 We base our study on the results of work\cite{Kuch3}. It
was shown that due to the spin-wave exchange there is an effective
long-range attraction between two holes with opposite spins
\begin{equation}
\label{pot}
U_{eff}(r) = \frac{\lambda}{r^2},\hspace{1.cm}\lambda < 0.
\end{equation}
In this potential there is an infinite series of two-hole bound
states. However they have very large sizes and very small
binding energies. Therefore these states cannot be directly
responsible for high-$T_c$ superconductivity.
In the present paper we demonstrate that in many-hole problem
the same potential (\ref{pot}) produces very strong pairing.

It is well known that the finite concentration of the holes
destroys the long-range antiferromagnetic order.
The starting point of our consideration is the assumption that a
local antiferromagnetic order is preserved. It is sufficient to suppose
the order to exist at least in the region of distances as large as the
wave length of a hole $\sim 1/p_F$. This assumption does not contradict
to experimental data (see e.g. Ref.\cite{Birg}). In conclusion of
the present paper we argue that the actual magnetic correlation
length is probably even larger.

We describe the pairing using the BCS-type wave function for the
ground state. The superconducting gap $\Delta_{\bf p}$ at
$|{\bf p}|=p_F$ is found to have the form
\begin{equation} \label{gap}
\Delta_{\bf p}=C~\mu~\exp(-1/g) \sin \varphi.
\end{equation}
Here $C$ is a constant, $C \sim 1$, $\mu$ is the chemical potential,
and $g$ is given by the expression
\begin{equation} \label{g}
g= \Lambda f^2 \rho {{2\sqrt{a}}\over{(\sqrt{a}+1)^2}},
\end{equation}
where $f$ is the constant of the hole--spin-wave interaction, $\rho =
\sqrt{m_1m_2}/\pi$ is the density of states which depends on two masses
describing the anisotropic hole dispersion near the minimum of the
band, $a$ is the mass ratio $a=m_2/m_1 \gg 1$, and the coefficient
$\Lambda>1$ appears due to the special ``collapse'' effect in the
attractive potential (\ref{pot}). The angle $\varphi$
marking the position of ${\bf p}$ on the Fermi surface is precisely
defined below.

The superconducting gap is presented in (\ref{gap}) in the form as
close as possible to the usual result in BCS theory. However there are
important differences which distinguish the problem under consideration
from the well-known phonon BCS theory and which lead to the very strong
pairing.\\
1)Due to the small concentration of the holes $\delta \ll 1$
the velocity of spin-wave $c$ is much higher then the velocity of
the hole: $c \gg v_F \propto \sqrt{\delta}$. Therefore the retardation
can be neglected and hole-hole interaction is described by the static
potential. As a result the nonexponential factor in
(\ref{gap}) is the chemical potential $\mu$ in contrast to the usual
situation where it is $\omega_D$.\\
2)There is a substantial enhancement of the superconducting gap caused
by the ``collapse'' effect. In the attractive potential (\ref{pot})
the wave function is known to collapse to the origin. Certainly due to
the finite lattice spacing there is no real collapse, but a trace of
this effect remains. This is the enhancement factor $\Lambda$ in the
formula (\ref{g}). This factor is a function of $p_Fr_0$, where
$r_0 \sim lattice \ spacing$ is a cutoff radius. We will demonstrate
that $\Lambda > 1$ for $p_Fr_0 \ll 1$.\\
3)An anisotropy of the hole dispersion ($a=m_2/m_1 \gg 1$)
plays an important role in the enhancement of the gap. Actually let
us fix $m_1$ and consider the effective coupling (\ref{g})
as a function of the mass ratio $a$. We see that $g(\infty)=4g(1)$.
For realistic values of the parameters of the $t$-$J$ model the
anisotropy factor equals $a \approx 7$, and $g(7) \approx 2g(1)$.
The general physical reason for this result is clear. The stronger is
the anisotropy the more our 2D system acquires the properties of a 1D
system. The interaction plays the more important role in 1D than in 2D.
Therefore the superconducting gap is to increase with the anisotropy.

 Formula (\ref{gap}) presents the gap on the Fermi surface
which is an ellipse situated on the face of the magnetic Brillouin
zone. The maxima of the absolute value of $\Delta_{\bf p}$ are in the
direction along the face of the zone, the two nodes are in the orthogonal
direction. In spite of this ``dipole'' nature, the parity of the state
with the gap (\ref{gap}) is positive. The state belongs either to the
symmetry which is often called $d$-wave or to the symmetry of $g$-wave.
In the present work we cannot distinguish between these two cases because
we do not take into account explicitly the transitions between different
pockets of the Fermi surface. However the detail numerical
calculation\cite{Bel4} demonstrates that $d$-wave symmetry is more
favorable.

In addition to (\ref{gap}) we find the series of the solutions of
BCS equation which behave as
$\Delta_{\bf p}^{(m)} \sim \sin m \varphi,~m=1,2,\ldots$.
However their absolute values sharply decrease with the increase of
$m$, that is why we pay the main attention to the case of $m=1$.
Let us note also that there is no solution for the gap without
nodes ($m=0$).

Our paper has the following structure.
In Section II we present an effective Hamiltonian describing the
long-range dynamics of the system. It is expressed in the terms of
dressed holes and spin-waves. This Section is based on the results
of Refs.\cite{Kuch3,Sus2,Suh3,Suhf,Cher3}.
In Section III we consider the BCS-type pairing in the weak coupling limit.
The ``collapse'' effect here is neglected: $\Lambda=1$.
Section IV presents the results of numerical solution of the BCS
equation which demonstrates that the gap is large and dependent on
cutoff parameter.
In Section V the ``collapse'' effect is considered.
The relation between superconducting critical temperature and the gap
is discussed in the Section VI.

%*****************************************************************
\section{Interaction between the holes with opposite spins. Effective
Hamiltonian for the dressed holes.}
%*****************************************************************
The $t$-$J$ model with less than half-filling is defined by the Hamiltonian
\begin{equation}
\label{H}
  H = H_t + H_J
    = -t \sum_{<nm>\sigma} ( d_{n\sigma}^{\dag} d_{m\sigma} + \mbox{H.c.} )
    + J \sum_{<nm>}  {\bf S}_n{\bf S}_m,
\end{equation}
where $d_{n\sigma}^{\dag}$ is the creation operator of a hole with spin
$\sigma$ ($\sigma= \uparrow, \downarrow$) at site $n$ on a
two-dimensional square lattice. The $d_{n\sigma}^{\dag}$ operator acts
in the Hilbert space where there is no double electron occupancy. The
spin operator is
${\bf S}_n = {1 \over 2} d_{n \alpha}^{\dag}
${\boldmath $\sigma$}$_{\alpha \beta} d_{n \beta}$.
$<nm>$ are neighbor sites on the lattice. Below, we set $J=1$.

At half-filling (one hole per site) the $t$-$J$ model is equivalent to
the Heisenberg antiferromagnet model which has the long-range
antiferromagnetic order in the ground state.
The problem is the behavior of the system under doping by additional
holes. It is well known that under the doping the long-range
antiferromagnetic order is destroyed. We believe that the short-range
antiferromagnetic order is preserved and the magnetic
correlation length $\xi_M$ is larger than all other scales which
are related to the pairing. Thus, we discuss all excitations of the
system starting from the antiferromagnetic background of the $t$-$J$
model. The quasiparticles are the dressed holes and spin-waves.

 Let us denote by $|0\rangle$ the wave function of quantum Neel
state with long-range antiferromagnetic order. The wave function
of a hole is of the form $\psi_{{\bf k}\sigma}=h^{\dag}_{{\bf k}\sigma}
|0\rangle $, where $h^{\dag}_{{\bf k}\sigma}$ is a creation
operator of a dressed hole, and $\sigma = \uparrow,\downarrow$.
An approximate expressions for $h^{\dag}_{{\bf k}\sigma}$ in terms
of bare hole operators $d^{\dag}_{{\bf k}\sigma}$ are presented in
Refs.\cite{Sus2,Suhf}. The dressed hole is a usual fermion.
We choose the spin-waves to have definite values of spin projection:
$\alpha^{\dag}_{\bf q}$ is a creation operator of spin-wave with
$S_z=-1$, and $\beta^{\dag}_{\bf q}$ is a creation operator of
spin-wave with $S_z=+1$ (see Ref.\cite{Manousakis} for the review
of spin-wave theory). The long-range dynamics of the system
is described by the effective Hamiltonian derived in the
Ref.\cite{Kuch3}
\begin{equation}
\label{Heff}
H_{eff}=\sum_{\bf k}\epsilon_{\bf k}h_{\bf k}^{\dag}h_{\bf k}+
\sum_{\bf q}\omega_{\bf q}(\alpha_{\bf q}^{\dag}\alpha_{\bf q}
+\beta_{\bf q}^{\dag}\beta_{\bf q}) + H_{h,sw} + H_{hh}.
\end{equation}
The first two terms in (\ref{Heff}) describe the free holes and
spin-waves, the other two account for their interaction. It is
well known (see e.g. Refs.\cite{Dag0,Mart1,Liu2,Sus1,Sus2,Giam3})
that the minima of single hole energy band are situated at the points
${\bf k}_0=(\pm \pi/2, \pm \pi/2)$ on the face of the magnetic
Brillouin zone, see Fig.1. Near the bottom the energy
$\epsilon_{\bf k}$ may be presented in the usual quadratic form
\begin{equation}
\label{eps}
\epsilon_{\bf k} \approx {1\over2}\beta_1 p_1^2+{1\over2}\beta_2 p_2^2,
\hspace{1.cm}{\bf p}={\bf k}-{\bf k}_0,
\end{equation}
where indices 1 and 2 label the projections of the vector onto the
directions orthogonal and parallel to the face of the magnetic Brillouin
zone, respectively (see Fig.1). Due to the
Refs.\cite{Dag0,Mart1,Liu2,Sus1,Sus2,Giam3} the inverse masses
$\beta_i=1/m_i$ can be approximated by the following formulae
\begin{equation}
\label{bet1}
\beta_1 \approx \left\{ \begin{array}{ll}
 2.1t^2 & \mbox{for $t\ll 0.33,$}\\
 0.65t  & \mbox{for $0.33 \ll t \le 5,$}
                        \end{array}
                \right.
\hspace{1.cm} a={{\beta_1}\over{\beta_2}} \approx 5-7.
\end{equation}
An approximation for $\beta_1$ valid for any $t \le 5$ is presented in
Ref.\cite{Sus2}.

  The spin-wave dispersion is known to be\cite{Manousakis}
\begin{equation} \label{om}
\omega_{\bf q}=2\sqrt{1-\gamma_{\bf q}^2} \to
\sqrt{2}|{\bf q}|, \quad {\rm at} \quad q \ll 1,
\end{equation}
where $\gamma_{\bf q}= \frac{1}{2}(\cos q_x+ \cos q_y)$. In this units
the velocity of spin-wave is $c=\sqrt{2}$.

The third term in (\ref{Heff}) which describes the interaction of
a composite hole with spin-wave is of the form (see, e.g.
Refs.\cite{Mart1,Liu2,Suh3,Suhf})
\begin{equation} \label{Hh}
H_{h,sw}= \sum_{{\bf k},{\bf q}}g({\bf k},{\bf q})
\biggl(h_{{\bf k}+{\bf q}\downarrow}^{\dag}
h_{{\bf k}\uparrow} \alpha_{\bf q}+h_{{\bf k}+{\bf q}\uparrow}^{\dag}
h_{{\bf k}\downarrow} \beta_{\bf q} + H.c. \biggr),
\end{equation}
where the vertex $g({\bf k},{\bf q})$ is
\begin{eqnarray} \label{gkq}
g({\bf k},{\bf q})&=&2\sqrt{2}f(\gamma_{\bf k}U_{\bf q}+
\gamma_{{\bf k}+{\bf q}}V_{\bf q}),\\
U_{\bf q} & = & \sqrt{\frac{1}{\omega_{\bf q}}+ \frac{1}{2}}, \nonumber\\
V_{\bf q} & = & - {\rm sgn}(\gamma_{\bf q}) \sqrt{\frac{1}{\omega_{\bf
q}}- \frac{1}{2}}.\nonumber
\end{eqnarray}
Here $f$ is a
hole-spin-wave coupling constant. In the perturbation theory limit
$(t \ll 0.33)$ and at $zS \gg 1$ ($z$ is the number of neighbor sites)
the coupling constant is $f = 2t$ (see e.g. Refs.\cite{Mart1,Liu2}).
For an arbitrary $t$ and realistic value $zS=2$ the coupling
constant was calculated in the Refs.\cite{Suh3,Suhf}. The plot
of $f$ as a function of $t$ is presented in Fig.2. For small
$t$: $f \approx 3.4t$. For large $t$ the coupling constant is
$t$-independent: $f \approx 2.$
In the vicinity of the band bottom (${\bf k} \approx {\bf k}_0$)
and for a small transferred momentum $(q \ll 1)$ the vertex
(\ref{gkq}) is reduced to the form
\begin{equation} \label{gk0}
g({\bf k},{\bf q}) \approx 2^{3/4}f\frac{q_1}{\sqrt{q}}.
\end{equation}
We recall that $q_1$ is the projection of ${\bf q}$ onto the axis
orthogonal to the face of the Brillouin zone, see Fig.1.

The last term $H_{hh}$ in (\ref{Heff}) describes the contact
hole-hole interaction arising from exchange of many spin-waves.
It is discussed in the Ref.\cite{Kuch3}. In the present paper we
do not need the explicit expression for this interaction.

%%%%%%%%%%%%%%%%%%%%%%%%%%%%%%%
\section{BCS-type pairing induced by the spin-wave exchange.
The weak coupling limit}
%%%%%%%%%%%%%%%%%%%%%%%%%%%%%%%

For the considered small concentrations $\delta \ll 1$, the holes are
localized in the momentum space in the vicinity of the minima of the
band $k_0=(\pm \pi/2, \pm \pi/2)$. Consider the hole dynamics in one
pocket of the Brillouin zone, for example, near the point
$k_0=(\pi/2, \pi/2)$, see Fig.1.
Remember that we denote by ${\bf p}$ the
deviation of the momentum ${\bf k}$ from the band bottom ${\bf p}={\bf
k}-{\bf k}_0$, the axis 1 is normal to and the axis 2 is parallel to
the face of the zone. The hole states with negative $p_1$ are outside
the Brillouin zone, but they are equivalent to those states inside the
zone which are located in the vicinity of the point $(- \pi/2,-
\pi/2)$. The equivalence follows from the simple relationship ${\bf
k}'={\bf k}-{\bf g}$, where ${\bf g}=(\pi, \pi)$ is the vector of the
inverse magnetic lattice.

Let us do some useful estimations. If we forget about the hole-hole
interaction we can easily calculate the Fermi energy and Fermi momentum
which is actually the deviation of momentum from the center of the pocket.
\begin{equation}
\label{ep}
\epsilon_F= \frac{\pi}{2}(\beta_1 \beta_2)^{1/2} \delta, \qquad p_F
\sim (\pi \delta)^{1/2}.
\end{equation}
In pairing the exchange of spin-wave with the typical momentum $q \sim
p_F$ is important. Due to Eq.(\ref{om}) the energy of this quantum is
much higher than the typical energy of a pair
\begin{equation}
\label{omega}
\omega_q \sim p_F \sim (\pi \delta)^{1/2} \gg \epsilon_F \sim (\beta_1
\beta_2)^{1/2} \delta.
\end{equation}
The situation is quite similar to that for the two-hole bound state
problem \cite{Kuch3}, and different from the situation with the usual
phonon induced pairing where Debye's frequency is much lower than
the Fermi energy.

Let us find the interaction between two holes with $S_z=0$ caused by
soft spin-wave exchange. This process is described by the diagram in
Fig.3. Calculating it with the help of (\ref{Hh}) we find
\begin{equation} \label{vpp}
V_{\bf pp'}\approx-2 \frac{g^2({\bf k},{\bf q})}{(- \omega_{\bf q})}
\approx 4f^2\frac{q_1^2}{q^2}.
\end{equation}
The sign ``-'' before the expression takes into account the fact that
the spin-wave exchange makes the spin flip for both holes. Due to the
same reason the transferred momentum is the sum (not the difference)
${\bf q}= {\bf p}+ {\bf p}'$ of the hole momenta ${\bf p},{\bf p}'$.
The energy denominator $(- \omega_{\bf q})$ in (\ref{vpp}) takes into
account the energy of spin-wave, neglecting the energies of the
two holes which, according to (\ref{omega}), are low.
The expression (\ref{vpp}) describing the interaction between two holes
caused by spin-wave exchange was first evaluated in Ref.\cite{Kuch3}.
It only depends on the transferred momenta ${\bf q}$. Therefore it has a
potential-type form.

We use usual BCS wave function for the ground state of many-hole system:
\begin{equation} \label{Psi}
\Psi= \prod_{\bf p} (u_{\bf p}+v_{\bf p}h^{\dag}_{{\bf p}
\uparrow}h^{\dag}_{-{\bf p} \downarrow})|0 \rangle.
\end{equation}
Thus we suppose that all quasiparticles are in condensate.
For strong interaction the validity of this supposition is under
question because there is no parameter to justify it. We believe that
numerically wave function (\ref{Psi}) is good. Anyway one may
consider the wave function (\ref{Psi}) as a trial one in the
variational method. In this case the large gain in energy which
we get is a justification of wave function.

The gap $\Delta_{\bf p}$  corresponding to the wave function (\ref{Psi})
satisfies to conventional BCS equation.
\begin{equation}
\label{Delta}
\Delta_{\bf p}= -\frac{1}{2} \sum_{\bf p'} V_{\bf pp'}
\frac{\Delta_{\bf p'}}{E_{\bf p'}},
\end{equation}
where
$E_{\bf p}=\sqrt{\xi_{\bf p}^2+ \Delta_{\bf p}^2}$,
$\xi_{\bf p}=\epsilon_{\bf p}- \mu$.
A very important question is the symmetry of wave
function (\ref{Psi}) which is related to the symmetry of
superconducting gap. Equation (\ref{Delta}) for the gap is invariant
under transformation ${\bf p} \rightarrow -{\bf p},~ {\bf p'}
\rightarrow -{\bf p'}$. Therefore we can to classify $\Delta_{\bf p}$
by the quantum number ${\cal R}$:
\begin{equation} \label{R}
\Delta_{-{\bf p}}= {\cal R} \Delta_{\bf p},\qquad {\cal R}= \pm 1.
\end{equation}
Note that the transformation ${\bf p} \rightarrow -{\bf p}$ is not the
inversion because it is applied not to the momentum ${\bf k}$ but to
the deviation ${\bf p}$ of the momentum from ${\bf k}_0,~{\bf p}= {\bf
k}-{\bf k}_0$. The connection between parity ${\cal P}$ and the quantum
number ${\cal R}$ was discussed in detail in \cite{Kuch3}. The result
obtained remains valid for the problem under consideration:
\begin{equation} \label{P}
{\cal P}=-{\cal R}.
\end{equation}
Let us explain the meaning of (\ref{P}) using simple examples. Consider
the positive parity, ${\cal P}=1$, gap  $\Delta_{\bf k}
\propto (\cos k_x- \cos k_y)$ which is usually called $d$-wave. Let us
expand it near the center of the pocket ${\bf k}={\bf k}_0+{\bf p}$. We
get $\Delta_{\bf p} \propto p_2$ at $p \ll 1$. Therefore due to
(\ref{R}) ${\cal R}=-1$. As a function of the angle of ${\bf p}$ the
function $\Delta_{\bf p}$ has two nodes.
Similarly, the negative parity, ${\cal P}=-1$, gap $\Delta_{\bf k}
\propto \sin k_x$, which is usually called $p$-wave, has ${\cal R}=+1$.
It has no nodes when we rotate around the center of the pocket.
Certainly the above functions are not localized at small ${\bf p}$, and
therefore their classification by the quantum number ${\cal R}$ instead
of parity is unnatural. In the present work as well as in Ref.\cite{Kuch3}
we consider the states localized at small ${\bf p}$ and using of
${\cal R}$ is very convenient.

 It is convenient to perform
the rescaling $p_1 \to \sqrt{\beta_1}p_1$, $p_2 \to \sqrt{\beta_2}p_2$
to isotropic single-hole dispersion. In these coordinates the BCS
equation is of the same form (\ref{Delta}), but $\xi_{\bf
p}=p^2/2-\mu$, and potential (\ref{vpp})
looks as follows:
\begin{eqnarray}
\label{BCS}
&&V_{{\bf pp}'}= 2\pi F\sqrt{a} {{q_1^2}\over
{q_1^2+aq_2^2}}, \qquad {\bf q}={\bf p}+{\bf p}',\nonumber\\
&&F=2f^2/(\pi \beta_1), \quad a=\sqrt{\beta_1/ \beta_2}.
\end{eqnarray}
The plot of constant $F$ as a function of $t$ is presented in Fig.2.
The value of $t$ corresponding to realistic superconductors is equal
$t \approx 2-3$ (see e.g. Refs.\cite{Esk0,Fla1,BCh3}), and therefore
$F \approx 1.0-1.2$. Let us note that hole-spin-wave coupling constant
$f$ is a result of renormalization\cite{Suhf} and therefore it
depends on typical momentum transfer $q$. The plots presented in Fig.2
corresponds to $q=0$. Actually we are interested in $q \sim p_F$.
Corresponding value of $f$ is about 10\% higher for hole concentration
$\delta \sim 0.1$. Therefore reasonable region for $F$ is
$F \approx 1.1-1.5$. Further we will call these values the
``physical values'' of $F$.

  At small $F$ the gap is small and localized in the vicinity of the
Fermi surface: $\Delta_{\bf p}=\Delta(\varphi)$. After integration
over $\xi$ with logarithmic accuracy and substitution
$|{\bf p}|=|{\bf p'}|=p_F$ into $V_{{\bf p}{\bf p'}}$ the equation
(\ref{Delta}) is transformed into
\begin{equation}
\label{BCS1}
\Delta(\varphi)=\ln\biggl({{C \mu}\over{\Delta}}\biggr)
\int_0^{2 \pi}K(\varphi + \varphi') \Delta(\varphi^{\prime})
\frac{d \varphi^{\prime}}{2\pi},
\end{equation}
where $\Delta=max|\Delta(\varphi)|$; $C$ is a number, $C\sim 1$;
and the kernel $K(\varphi)$ is
\begin{equation}
\label{K}
K(\varphi) =
-F\sqrt{a}{{1+\cos \varphi}\over{a+1-(a-1) \cos \varphi}}.
\end{equation}
Eq.(\ref{BCS1}) is closely related to the eigenvalue problem for
the kernel $\hat K$:
\begin{equation}
\label{K1}
\frac{1}{2 \pi}~\int_0^{2\pi}K({\varphi}+{\varphi}^{\prime})
\psi(\varphi^{\prime}){d \varphi^{\prime}}=g\psi(\varphi),
\end{equation}
where $g$ is an eigenvalue.
One can easily check that the solutions of this equation are
\begin{eqnarray}
\label{EK}
&&g_m={{Fa}\over{a-1}}\biggl({{\sqrt{a}-1}\over{\sqrt{a}+1}}\biggr)^m,
\nonumber\\
&&\psi_m(\varphi)=\frac{1}{\sqrt{\pi}}\sin m \varphi,
\end{eqnarray}
where $m=1,2,3, \ldots$.
Comparing (\ref{K1}) with (\ref{BCS1}) and using (\ref{EK}) we find the
series of solutions for the superconducting gap on the
Fermi surface in the weak coupling limit:
\begin{equation}
\label{ga}
\Delta_m(\varphi)=C \mu \exp(-1/g_m)\sin m \varphi.
\end{equation}
All the solutions found possess nodes on the Fermi surface. From
(\ref{ga}) and (\ref{R}) we find that ${\cal R}=(-1)^m$ which gives the
parity
\begin{equation}
\label{P1}
{\cal P}=-(-1)^m.
\end{equation}
We are interested in the mass ratio $a \sim 5-7$ (\ref{bet1}). In spite
of this large mass ratio $a$ the value of $g_m$
(\ref{EK}) decreases very rapidly with the increase of $m$. Therefore
further we will consider mainly the case $m=1$.

Let us note that we should consider the solution with $m=1$ in each
pocket of the Brillouin zone. After that we have to take symmetrical or
antisymmetrical combination between the pockets. One of them belongs
to $B_1$ representation of lattice symmetry group $C_{4v}$, and
another combination belongs to $A_2$ representation. The $B_1$
solution has the same symmetry as the wave function usually called
$d$-wave $\Delta_{\bf k} \propto (\cos k_x - \cos k_y)$
(the $A_2$ solution has the symmetry of $g$-wave).
However we stress that except the symmetry there is nothing common
between the solution (\ref{ga}) and this wave function.

Thus in the present section we have derived the formulae
(\ref{gap}) and (\ref{g}) except the factor $\Lambda$ describing
the ``collapse effect''. The reason for this is clear: In the
weak-coupling limit there is no ``collapse'' effect. This effect
is considered below.

%*****************************************************************
\section {Numerical solution of BCS-equation}
%*****************************************************************

We will see that solution of the BCS equation (\ref{Delta})
at physical value of hole-spin-wave coupling
constant $f$ is divergent at large momenta. The origin of this
divergency is the same as one in the two-hole problem\cite{Kuch3}, that
is the collapse of the wave function. Physically there is no problem
with this divergency because of a finite value of lattice spacing.
Similarly to the two-hole problem we have to solve the BCS equation
(\ref{Delta}) numerically with both spin-wave (\ref{vpp}) and contact
interaction $H_{hh}$ taken into account, and with momentum restricted
within the Brillouin zone. Such a solution will be presented
elsewhere\cite{Bel4}.
In the present work we are interested mainly in qualitative understanding
of the problem and in the dependence of the solution on the parameters.
To elucidate it we introduce a cut-off factor into the interaction
(\ref{BCS}):
\begin{equation}
\label{Q}
V_{{\bf p}{\bf p}^{\prime}}^Q=
V_{{\bf p}{\bf p}^{\prime}}{{Q^2}
\over{{\bf q}^2+Q^2}}
\end{equation}
Numerically we set $Q$ to be of the order of the vector of inverse
lattice, $Q\sim 1$, but we will consider it as a parameter.

Numerical solution of equation (\ref{Delta}) with the interaction
(\ref{Q}) is performed in one pocket of the Brillouin zone, assuming
the quadratic dispersion (\ref{eps}) to be valid for any momentum
$p$. We set $t=3$, $\beta_1=1.95$ and $a=7$ in accordance with
(\ref{bet1}). The chemical potential is chosen to be
$\mu=5.77\cdot 10^{-2}$. This value corresponds to the concentration
of holes $\delta=0.05$ for ideal gas, see Eq.(\ref{ep}).
In accordance with previous section
the gap was found to be maximal in the sector corresponding to $m=1$,
where ${\cal R}=-1$. It has two nodes as a function of the
angle. Generally, the angular dependence of the gap is rather
complicated, but at the Fermi surface ($p=\sqrt{2\mu}$) it is
$\Delta_{\bf p}=\Delta_1 \sin{\varphi}$ with very good accuracy.
Due to the results of previous section it is nor surprising for
small $F$. At large $F$ the smallness of correction is due to the
specific  spectrum of operator $\hat K$ (\ref{K}). The explanation
of this observation is given in Appendix.

The plot of $\Delta_1/\mu \times \exp(1/g_1)$ is presented in
Fig.4 as a function of the coupling constant $F$. There
are two curves which differ only in the value of the cutoff
parameter: $Q=1.5$ and $Q=2$. We stress that at the physical
values of coupling constant ($F\approx 1.1-1.5$) the gap is very
large: $\Delta \sim \mu$.

Concentration of the holes corresponding to the wave function
(\ref{Psi}) is equal
\begin{equation}
\label{del}
\delta=2{{1}\over{\sqrt{\beta_1 \beta_2}}}\sum_{\bf p}2v_{\bf p}^2
\end{equation}
The factor $1/\sqrt{\beta_1 \beta_2}$ is due to the rescaling of
coordinates, and an additional factor 2 is due to the two pockets in
the Brillouin zone. In the same Fig.4 the curves of $\delta/0.05$ as a
function of $F$ are presented for $Q=1.5$ and $Q=2$.

We see from Fig.4 that 1)the ratio $\Delta_1/\mu \times \exp(1/g_1)$
depends on the coupling constant $F$, 2)It depends on the cut-off
parameter $Q$. The dependence on $F$ is not so surprising because
formula (\ref{ga}) is derived only with logarithmic accuracy. More
important is $Q$-dependence. This dependence is related to the
high-momentum ($p \gg p_F$) behavior of BCS equation (\ref{Delta}). At
such a momentum it is equivalent to the Schrodinger equation for the
two-hole problem. Indeed, introducing
$\psi_{\bf p}=\Delta_{\bf p}/E_{\bf p}$
one can rewrite (\ref{Delta}) as
\begin{equation}
\label{Sch}
-{1\over 2}\sum_{\bf p^{\prime}}V_{{\bf p}{\bf p}^{\prime}}
\psi_{\bf p^{\prime}}=
E_{\bf p} \psi_{\bf p}
\approx (\frac{{\bf p}^2}{2}-\mu) \psi_{\bf p}.
\end{equation}
The last transformation is valid only at $p \gg p_F$. Note that in the
usual phonon picture this limit is forbidden because the momentum is
limited $p \sim p_F$ due to the smallness of Debye's frequency. Thus,
to understand the origin of $Q$-dependence one
has to come back to the two-hole problem. It was shown in \cite{Kuch3}
that in the coordinate representation the variables in the Schrodinger
equation are separated and the effective potential for the radial
motion is of the form (\ref{pot}). In the sector $m=1$ the potential
parameter
$\lambda$ is positive for small $F$. When $F$ exceeds some critical
value, $\lambda$ becomes negative and singular attraction arises.
According to Ref.\cite{Kuch3} for $a=7$ the critical value is $F_c
\approx 0.67$ and this gives us a qualitative explanation of the
$Q$-dependence.

Let us look at the plots of hole concentrations as functions of $F$ at
a fixed chemical potential which are presented in Fig.4.
At small $F$ the density coincides with that of the ideal gas. However
it increases with $F$. It means that if we keep the density fixed the
chemical potential drops with $F$. It was shown in Ref.\cite{Kuch3}
that the potential (\ref{pot}) for negative $\lambda$ gives a two-hole
bound state of the size
\begin{equation}
\label{size}
r \sim \exp\biggl({{\pi}\over{\sqrt{|\lambda|}}}\biggr)
\end{equation}
For $F=1.1-1.5$ and $a=7$ the value of $\lambda$ is
$\lambda = -1--2$ and $r \sim 10-30$. Therefore at a realistic density
the separation between the holes is smaller than the bound state size.
However, if we increase $F$ at a fixed density, the bound-state size
decreases and
finally we get the Bose condensate of two-hole ``atoms'' of a size
smaller than the separation between the ``atoms''. Due to
Eq.(\ref{Sch}) the chemical potential $\mu$ in this case is negative
and equal to half of the energy of the bound state. The plots of
$\delta$ presented in Fig.4 reflect the tendency of chemical potential
to approach this negative value starting from the initial positive
value corresponding to the ideal gas. From this consideration it is
obvious that at $F\approx 1.1-1.5$, but at very low density the ground
state BCS wave function (\ref{Psi}) is just the Bose condensate of
the ``atoms''.

Thus in the present section  we have got the numerical solution of BCS
equation at an arbitrary coupling constant $F$.  We have demonstrated
also that at the physical values of the coupling constant ($F\approx
1.1-1.5$) the gap is large, $\Delta \sim \mu$.

%%%%%%%%%%%%%%%%%%%%%%%%%%%%%%%
\section {The collapse effect}
%%%%%%%%%%%%%%%%%%%%%%%%%%%%%%%
In this section  we consider the case of coupling strong enough to
produce the effect of collapse of the wave function in the region of
small distances $r \sim 1$. At first let us rewrite the BCS equation
in a form convenient for the analysis of this effect.
Consider the equation (\ref{Delta}) in isotropic coordinates
($\epsilon_{\bf p}=p^2/2$). Let us introduce a new variable
$\Delta_{\bf p} \to \chi_{\bf p}$
\begin{equation}
\label{hi}
\chi_{\bf p}={{\Delta_{\bf p}}\over
{\epsilon_{\bf p}+\mu}}.
\end{equation}
Then ({\ref{Delta}) is transformed to
\begin{equation}
\label{f1}
(\epsilon_{\bf p}+ \mu) \chi_{\bf p}=
-{1\over 2} \sum_{\bf p^{\prime}} V_{{\bf p}{\bf p}^{\prime}}
\frac{\epsilon_{{\bf p}'}+\mu}{E_{\bf p^{\prime}}} \chi_{{\bf p}'},
\end{equation}
and further
\begin{equation}
\label{f2}
\left(\epsilon_{\bf p}+{1\over{2}}{\hat V}+ \mu \right) \chi_{\bf p}=
-{1\over 2}\sum_{\bf p^{\prime}}V_{{\bf p}{\bf p}^{\prime}}
\left( \frac{\epsilon_{\bf p^{\prime}}+\mu}{E_{\bf p^{\prime}}}-1 \right)
{1\over{\epsilon_{\bf p^{\prime}}+ \mu}}( \epsilon_{\bf p^{\prime}}+
\mu) \chi_{\bf p^{\prime}}.
\end{equation}
The gap $\Delta_{\bf p}=(\epsilon_{\bf p}+\mu) \chi_{\bf p}$ is
not divergent at large $(p \gg p_F)$ momenta. Therefore the
integrand in Eq.(\ref{f2}) is localized near $\mu$. The gap is a
smooth function of $\epsilon$. Therefore at $\Delta \ll \mu$ the
integral over $\xi$ in right hand side of Eq.(\ref{f2}) can be
calculated with logarithmic accuracy:
\begin{equation}
\label{f3}
\left( \frac{\epsilon_{{\bf p}'}+\mu}{E_{{\bf p}'}}-1
\right){1\over{\epsilon_{{\bf p}'}+ \mu}} \to 2L
\delta(\xi_{{\bf p}'}),
\end{equation}
where $\delta(\xi)$ is $\delta$-function, and
\begin{equation}
\label{f4}
L={1\over2} \int_0^\infty \left( \frac{\epsilon_{\bf
p}+\mu}{E_{\bf p}}-1 \right){{d \xi_{\bf p}}\over{\epsilon_{\bf
p}+ \mu}} \approx \ln \left( \frac{C \mu}{\Delta}
\right).
\end{equation}
Here $\Delta=max \Delta_{\bf p}|_{\epsilon_{\bf p}= \mu}$, and
$C \sim 1$ is a constant.
Thus at $\Delta \ll \mu$ the BCS equation is transformed to the
form
\begin{equation} \label{f7}
\left( p^2+ \hat V+ 2 \mu \right) \chi_{\bf p}=- \frac{L \mu}{\pi^2}
\oint V_{{\bf pp}'} \chi_{{\bf p}'} d \varphi',
\end{equation}
or equivalently
\begin{equation} \label{f8}
\chi_{\bf p}=-{{L \mu}\over{\pi^2}} \oint (GV)_{{\bf pp}'}
\chi_{{\bf p}'}~d\varphi',
\end{equation}
where $G$ is Green's function of the two-hole problem at energy
$E=-2\mu$
\begin{equation} \label{h2}
G={{1}\over{p^2+\hat V+2\mu}}.
\end{equation}

The momentum ${\bf p}'$ in (\ref{f8}) is on the Fermi surface, ${\bf
p}$ is arbitrary. This is very advantageous, because putting ${\bf p}$
on the Fermi surface one can consider (\ref{f8}) as an eigenvalue
problem on the Fermi surface. Solving it we find $L$ and the gap on the
Fermi surface. Then from the same equation (\ref{f8}) one can find
$\chi_{\bf p}$ anywhere, and therefore $\Delta_{\bf p}$, which is
related to $\chi_{\bf p}$ by (\ref{hi}).

Now we are going to set $|{\bf p}|=p_F$ and solve the eigenvalue problem
(\ref{f8}). First of all let us note that the results of the
weak-coupling limit could easily be evaluated from (\ref{f8}). In this
limit one can neglect the interaction $\hat V$ in Green's function.
Then in Eq.(\ref{f8}) $G \to 1/4\mu$, and this equation is reduced to
Eq.(\ref{BCS1}) reproducing the results of Section  III.

Now we want to go further and to do calculations with the exact Green's
function. We know from Ref.\cite{Kuch3} that in coordinate representation
the potential $\hat V$ is a local function. Moreover, in this
representation the variables in the Schrodinger equation are separated.
Let us use this advantage of coordinate representation and write
Eq.(\ref{f8}) in the form
\begin{equation} \label{fa}
\chi_{\bf p}=- \frac{L \mu}{\pi^2} \int \biggl(
d^2{\bf R}e^{-i{\bf p}{\bf R}}G({\bf R},{\bf r})V({\bf r})
e^{i{\bf p^{\prime}}{\bf r}}d^2{\bf r}\biggr)
\chi_{\bf p^{\prime}}~d\varphi^{\prime}.
\end{equation}
We recall that the angular dependence of $\chi_{\bf p}$ is very
close to $\sin{\varphi}$, see Appendix. Substituting the anzats
$\chi_{\bf p}=\chi \sin{\varphi}$ into Eq.(\ref{fa}) we find after
integration over $\varphi$ and $\varphi^{\prime}$
\begin{equation} \label{fi}
{1\over{L}}=-{{4\mu}\over{\pi}}\int \sin({\varphi_R})J_1(pR)
G({\bf R},{\bf r})V({\bf r})\sin({\varphi_r})J_1(pr)
d^2{\bf R}d^2{\bf r}.
\end{equation}
Here $J_1(x)$ is the Bessel function, $p=\sqrt{2\mu}=\sqrt{2}p_F$,
The fact that the variables in the two-hole Schrodinger equation are
separated means that there is a complete set of angular eigenstates:
$1=\sum_m |\Phi_m\rangle\langle \Phi_m|$. We can transform
the integrand in Eq.(\ref{fi}):
\begin{equation} \label{f9}
\int \sin(\varphi_R)G \sum_m |\Phi_m\rangle\langle \Phi_m|V
\sin(\varphi_r)d\varphi_R d\varphi_r
=\sum_m \langle\sin(\varphi_R)|\Phi_m\rangle G
\langle \Phi_m|V|\sin(\varphi_r)\rangle
\end{equation}
At $F=0$ we have $V=0$, and the angular eigenstates are of the form
$\Phi_m={1\over{\sqrt{\pi}}}\sin m\varphi$. With increasing $F$ they
are deformed, but due to the Ref.\cite{Kuch3} $\Phi_1$ is very close
to ${1\over{\sqrt{\pi}}}\sin \varphi$ even at rather high $F$.
The reason is obvious: The mixing of $\sin \varphi$ with
$\sin m\varphi$ ($m \ge 3$) is suppressed by the factor $1/(m^2-1)$
arising from denominator of perturbation theory.
Thus we can saturate the sum in
(\ref{f9}) by one term:
\begin{equation} \label{f10}
\langle\sin(\varphi_R)|\Phi_1\rangle G
\langle \Phi_1|V|\sin(\varphi_r)\rangle
\approx G\int V({\bf r})\sin^2\varphi_r d\varphi_r.
\end{equation}
Using the explicit expression for $V({\bf r})$ from Ref.\cite{Kuch3}
one can easily calculate the integral in (\ref{f10})
\begin{equation} \label{f11}
\int V({\bf r})\sin^2\varphi d\varphi=-{1\over{r^2}}Fa
\int_0^{2\pi}{{\sin^2\varphi-a\cos^2\varphi}
\over{(\sin^2\varphi+a\cos^2\varphi)^2}}\sin^2\varphi d\varphi=
-{{2\pi}\over{r^2}}g_1,
\end{equation}
where $g_1$ is given by formula (\ref{EK}) at $m=1$. After these
transformations we reduce equation (\ref{fi}) to the form
\begin{equation}
\label{f12}
{1\over{L}}=8\mu g_1 \int G(R,r)J_1(pR)J_1(pr){1\over{r^2}}dRdr.
\end{equation}
Here $G(R,r)$ is the radial Green's function of Schrodinger equation
at energy $E=-2\mu$.
We conclude from Eq.(\ref{f12}) that accounting for exact Green's
function gives the expression
\begin{equation}
\label{f14}
\Delta_{\bf p} =C \mu \exp \left(-1/(\Lambda g_1) \right) \sin \varphi,
\end{equation}
for the gap on the Fermi surface (cf. with Eq.(\ref{gap})).
The factor $\Lambda$ is equal to
\begin{equation}
\label{f15}
\Lambda=8\mu \int G(R,r)J_1(pR)J_1(pr){1\over{r^2}}dRdr.
\end{equation}
The radial Green's function $G(R,r)$ in (\ref{f15}) is easily
calculated using the representation
\begin{equation}
\label{f13}
G(R,r)={{f_0(r_{<})f_{\infty}(r_{>})}\over{w}}.
\end{equation}
Here $f_0(r)$ is the solution of radial Schrodinger equation regular at
small $r$, $f_\infty(r)$ is that regular at $r = \infty$. The factor
$w$ in (\ref{f13}) is the wronskian of these solutions.

At $F \to 0$ $\Lambda \to 1$. However the most interesting case is
``collapse'' one: $\lambda \le 0$ ($F \ge F_c$). In this case the potential
(\ref{pot}) gives the ``collapse'' of the wave function to the center. Let us
introduce the hard core of the radius $r_0$ which gives us ultraviolet
cutoff. The solutions of the Schrodinger equation are of the form
\begin{eqnarray}
\label{f18}
&&f_0(r)=\sqrt{r}\biggl({\rm Re}~ I_{\nu}(pr)+
bK_{\nu}(pr)\biggr),\\
&&f_{\infty}(r)=\sqrt{r}K_{\nu}(pr). \nonumber
\end{eqnarray}
Here $I_{\nu}(x)$ is modified Bessel function of the first kind and
$K_{\nu}(x)$ is modified Bessel function of the second kind;
 $\nu=\sqrt{\lambda}=i \alpha,~\alpha$ is real. The coefficient
$b$ should be found from the condition of vanishing of the  wave
function at $r=r_0$:
\begin{equation}
\label{f19}
b(x_0)=- \frac{{\rm Re}~I_\nu(x_0)}{K_\nu(x_0)}
\approx i{{\sinh(\pi \alpha)}\over{\pi}}\biggl(
{{{{(x_0/2)^{i\alpha}}/{\Gamma(1+i\alpha)}}+
{{(x_0/2)^{-i\alpha}}/{\Gamma(1-i\alpha)}}}\over
{{{(x_0/2)^{i\alpha}}/{\Gamma(1+i\alpha)}}-
{{(x_0/2)^{-i\alpha}}/{\Gamma(1-i\alpha)}}}}\biggr).
\end{equation}
Here $x_0=\sqrt{2 \mu}r_0$, and $\Gamma(z)$ is Euler's gamma function.
Remember that the physical condition is that the average separation
between the holes is smaller than the size of two-hole bound state.
Therefore $\alpha \ln(1/x_0)<1$ and $b(x_0)$ is a monotonic function.
Substituting (\ref{f18}) into (\ref{f13}),(\ref{f15}) we
get the final formula for $\Lambda$:
\begin{equation}
\label{f20}
\Lambda(x_0)=\Lambda_1+ b(x_0) \Lambda_2,
\end{equation}
where
\begin{eqnarray} \label{f20a}
\Lambda_1&=&4\int_0^{\infty} dx_2 \int_0^{x_2} dx_1
J_1(x_1)J_1(x_2)~{\rm Re}~I_\nu(x_1)K_\nu(x_2) \left( \frac{x_1}{x_2}+
\frac{x_2}{x_1} \right), \nonumber\\
\Lambda_2&=&4 \int_0^{\infty}J_1(x)K_{\nu}(x)xdx
\int_0^{\infty}J_1(y)K_{\nu}(y){1\over{y}}dy. \nonumber
\end{eqnarray}
The integrals are convergent at $r=0$. Therefore we set the lower
limits $r=0$ instead of $r=r_0$. Formula (\ref{f20}) gives the
dependence of the gap on $x_0=\sqrt{2 \mu}r_0$ and $\lambda$. At the
critical point $\lambda=0$ ($F=F_c$) formula (\ref{f20}) is reduced to
the simple form:
\begin{equation} \label{f21}
\Lambda(x_0)=1.71-{{1.01}\over{\ln(1.12/ x_0)}},
\end{equation}

The results of calculation of the gap based on (\ref{f14}), (\ref{f20})
are in qualitative agreement with the results of direct numerical solution
of the BCS equation presented in Fig.4. There is no quantitative
agreement because of A)formula (\ref{f14}) has been derived only
for $\ln(\mu/\Delta) \gg 1$, B)the different procedures of cut-off
have been used for numerical and analytical calculations.

In the present section we have demonstrated that due to the singular
nature of the potential (\ref{pot}) the gap on Fermi surface is
enhanced. This effect becomes very substantial when $F$ exceeds the
critical value, $F>F_c$. Below the critical value, $F<F_c$, the gap
still depends on the ultraviolet cut-off governing the behavior of
the wave function at small distances. The dependence disappears
at $F \ll F_c$. We stress that physical values of $F=1.1-1.5$ are
larger than $F_c \approx 0.67$.

%*****************************************************************
\section {Relation between critical temperature and gap at zero
temperature}
%*****************************************************************

Here we would like to discuss a connection between the gap
$\Delta_{\bf p}$ near the Fermi surface at zero temperature and critical
temperature of superconductivity $T_c$. We suppose that
the short-range antiferromagnetic order at distances $\sim 1/p_F$
survives near the critical temperature. Moreover, we assume in this
section that the finite size of the magnetic correlation length $\xi_M$
impose no restriction on the critical temperature. The validity of this
assumption has to be checked in a separate work.

The BCS equation for the critical temperature $T_c$ looks as follows:
\begin{equation}
\label {V1}
\Delta_{\bf p}= - \frac{1}{2} \sum_{\bf p'} V_{\bf pp'}
\Delta_{\bf p'}\frac{1}{\xi_{\bf p'}} \tanh
\frac{\xi_{\bf p'}}{2 T_c}.
\end{equation}
Taking into account that at $|{\bf p}|=p_F$ the relation
$\Delta_{\bf p} \approx \Delta _1 \sin\varphi$ is valid, and calculating
the integrals over $d\xi$ in Eqs.(\ref{Delta}),(\ref{V1}) with logarithmic
accuracy we get
\begin{equation}
\label{V4}
2 \ln\biggl(\frac{\pi}{\gamma} \frac{T_c}{\Delta_1(T=0)}\biggr)
\approx \frac{1}{\pi} \int_0^{2\pi} \sin^2 \varphi \ln (\sin^2 \phi)

q
q
q
\end{equation}
Therefore
\begin{equation}
\label{V5}
T_c \approx \frac{\gamma\sqrt{e}}{2\pi}\Delta_1(T=0)
\approx 0.5\Delta_1(T=0).
\end{equation}
Here $\ln\gamma=0.577$ is Euler constant.

The relation (\ref{V5}) connects the critical
temperature with the maximal value of the gap on the Fermi surface at
the same values of the chemical potential $\mu$ rather than at the same
values of the hole concentration. However, in two-dimensional case
dependence of the chemical potential $\mu$ on the temperature $T$ is
very weak. For the free electron gas it is given by the equation
\begin{equation}
\label {V6}
\mu(0)= \mu+T \ln(1+ \exp(- \mu/T)).
\end{equation}
Thus, at $T_c< \mu$ the difference between $\mu(T_c)$ and $\mu(0)$ is
small and can be neglected. The direct numerical solution of
Eqs.(\ref{Delta}) and (\ref{V1}) confirms the relation (\ref{V5}).

%*****************************************************************
\section {Conclusion}
%*****************************************************************
We demonstrate that the spin-wave exchange results in very strong
pairing in the ensemble of holes. We use the BCS wave function to
describe this pairing and find the gap to be very large, of the order
of chemical potential, $\Delta \sim \mu$. This fact permits one to hope
that the explanation of high-$T_c$ superconductivity may be found in
the direction of research suggested in this paper.

There are several very important points which are to be addressed next.\\
1. In the present work we consider the problem only in one pocket of
Brillouin zone with effective ultraviolet cutoff. The detail
numerical calculation for realistic Brillouin zone and with contact
hole-hole interaction taken into account will be presented in the
Ref.\cite{Bel4}.\\
2. Calculating the potential arising due to spin-wave exchange we do
not consider the renormalization of the spin-wave Green function caused
by the spin-polarization of the hole liquid. This approximation is
justified by inequality (\ref{omega}):
$\omega_q \sim p_F \sim (\pi \delta)^{1/2} \gg \epsilon_F \sim
(\beta_1\beta_2)^{1/2} \delta$. However for realistic density the
parameter is not so good. Account for Green function renormalization
enhances the interaction. That is why one can expect even stronger pairing
than the one predicted in this paper.\\
3. The most challenging is the problem of destruction by the doping
of the long-range antiferromagnetic order. The first
step on this way is calculation of renormalized spin-wave Green
function for small momenta ($q \ll p_F$). For normal Fermi liquid of the
holes it has been done in the work\cite{SF}. Now one has to do it for
correlated hole liquid. The first estimations show that similar to
normal Fermi liquid this renormalization is strong enough to destroy the
long-range antiferromagnetic order.  The naive estimation for magnetic
correlation length $\xi_M \sim 1/q_M$ follows from condition of intersection
of spin-wave spectrum with particle-hole continuum:
$\omega_{q_M} \sim \Delta$. Since $\omega_{q_M} \sim q_M$ and
$\Delta \sim \mu \sim \delta$ one gets $\xi_M \sim 1/\delta$.
Probably more correct estimation is $1/\delta > \xi_M > 1/p_F\sim
1/\sqrt{\delta}$. This question will be considered elsewhere.

{\bf ACKNOWLEDGMENTS}

We are grateful to J.Oitmaa for the discussion of Coulomb repulsion
of the holes. We are also very grateful to L.S.Kuchieva for the help in
preparing the manuscript.
%%%%%%%%%%%%%%%%%%%%%%%
\appendix
\section {The gap angular dependence on the Fermi surface.}
%%%%%%%%%%%%%%%%%%%%%%%

Due to the Section III for $m=1$ and for $\Delta_{\bf p} \ll \mu$
the angular dependence of a gap on the Fermi surface $(p=\sqrt{2\mu})$
is of the form $\Delta_{\bf p} \propto \sin\varphi$. Now we are going
to demonstrate that the correction to this formula is very small and
therefore it is practically valid even for $\Delta \sim \mu$.
The parameter which additionally suppresses this correction is
$(\sqrt{a}-1)^2/(\sqrt{a}+1)^2 \sim 0.2$.

The equation (\ref{BCS1}) was derived from BCS equation (\ref{Delta})
with logarithmic accuracy. Let us take into account first nonlogarithmic
correction. It is obvious that with this correction the Eq.(\ref{BCS1})
is modified to
\begin{equation}
\label{A1}
\Delta (\varphi)= \frac{1}{2 \pi} \int_0^{2 \pi}
K(\varphi+\varphi')L(\varphi') \Delta(\varphi') d \varphi',
\end{equation}
where $L(\varphi)$ is a sum
\begin{equation}
\label{A2}
L(\varphi)=L+l(\varphi)
\end{equation}
of a big logarithm $L=\ln (C\mu/\Delta)$ and an angle-dependent
correction $l(\varphi)$. Without $l(\varphi)$ Eq.(\ref{A1}) coincides
with the eigenvalue problem (\ref{K1}), whose spectrum is found in
(\ref{EK}). Let us use this spectrum in order to calculate the
correction caused by $l(\varphi)$. Using the conventional perturbation
theory one finds
\begin{eqnarray}
\label{A3}
&&\delta \Delta(\varphi)= \Delta_1 \sum_m
\frac{l_{m,1}}{\frac{1}{g_m}-
\frac{1}{g_1}} \sin m\varphi.\\
&&l_{m,1}=\frac{1}{\pi}\int_0^{2\pi} \sin(m\varphi)
l(\varphi) \sin(\varphi) d \varphi. \nonumber
\end{eqnarray}
Note that here $m=3,5,\ldots$.

The point is that there appears large number $1/g_m$ in the
denominator. For example, for $F=1$ and $a=7$ $g_3$ defined in
(\ref{EK}) is $g_3=0.1$. That is why the correction (\ref{A3}) is small.

\newpage
{\bf FIGURE CAPTIONS}

FIG. 1. The Brillouin zone of a hole in the $t-J$ model.\\

FIG. 2. The coupling constants. Solid line represents the
hole-spin-wave coupling constant $f$ calculated in Ref.\cite{Suhf}.
Dashed line represents the effective constant $F=2f^2/(\pi\beta_1)$
which governs the long-range interaction between the holes,
see Eq.(\ref{BCS}).\\

FIG. 3. Single spin-wave exchange between two holes.\\

FIG. 4. The results of numerical solution of BCS equation as a
function of $F$, for $Q=1.5$ and $Q=2.0$.\\
Solid lines present the ratio of maximal value of the gap on the Fermi
surface $\Delta_1$ to the expectation of that in weak-coupling limit:
$\Delta_1/\mu \times \exp(1/g_1)$. \\
The dashed lines present the ratio of density to that of ideal gas:
$\delta/0.05$.
\end{document}